\documentclass[runningheads]{llncs}
\usepackage[T1]{fontenc}
\usepackage{graphicx}
\usepackage{amsfonts}
\usepackage{booktabs}
\usepackage{tabularx}
\usepackage{gensymb}
\usepackage{makecell}
\usepackage{caption}
\usepackage{subfigure}
\usepackage{float}
\usepackage{marvosym}  % corresponding author
\usepackage{hyperref}

\begin{document}
\title{NAF: Neural Attenuation Fields for Sparse-View CBCT Reconstruction}
%
%\titlerunning{Abbreviated paper title}
% If the paper title is too long for the running head, you can set
% an abbreviated paper title here
%
\author{Ruyi Zha\textsuperscript{(\Letter)}\and Yanhao Zhang \and Hongdong Li}
% index{Last Name, First Name}
% index{Zha, Ruyi \and Zhang, Yanhao \and Li, Hongdong}
\authorrunning{R. Zha \textit{et al.}}
% First names are abbreviated in the running head.
% If there are more than two authors, '\textit{et al.}' is used.
%
\institute{Australian National University, Canberra, Australia
% \and   ***, ***, ***
%\email{***@***.com}\\
%\url{http://www.springer.com/gp/computer-science/lncs} \and
%ABC Institute, Rupert-Karls-University Heidelberg, Heidelberg, Germany\\
\\
\email{ruyi.zha@anu.edu.au}}
\maketitle              % typeset the header of the contribution
\begin{abstract}
This paper proposes a novel and fast self-supervised solution for sparse-view  CBCT reconstruction (Cone Beam Computed Tomography) that requires no external training data. Specifically, the desired attenuation coefficients are represented as a continuous function of 3D spatial coordinates, parameterized by a fully-connected deep neural network. We synthesize projections discretely and train the network by minimizing the error between real and synthesized projections. A learning-based encoder entailing hash coding is adopted to help the network capture high-frequency details. This encoder outperforms the commonly used frequency-domain encoder in terms of having higher performance and efficiency, because it exploits the smoothness and sparsity of human organs. Experiments have been conducted on both human organ and phantom datasets. The proposed method achieves state-of-the-art accuracy and spends reasonably short computation time. Code available at \url{https://github.com/Ruyi-Zha/naf_cbct}.

\keywords{CBCT \and Sparse View\and Implicit Neural Representation}
\end{abstract}

%
%
%%%%%%%%%%%%%%%%%%%%%%%%%
% Introduction
%%%%%%%%%%%%%%%%%%%%%%%%%
\section{Introduction}
Cone Beam Computed Tomography (CBCT) is an emerging medical imaging technique to examine the internal structure of a subject noninvasively. A CBCT scanner emits cone-shaped X-ray beams and captures 2D projections at equal angular intervals. Compared with the conventional Fan Beam CT (FBCT), CBCT enjoys the benefits of high spatial resolution and fast scanning speed~\cite{scarfe2006clinical}. Recent years have witnessed the blossoming of low dose CT, which delivers a significantly lower radiation dose during the scanning process. There are two ways to reduce the dose: decreasing source intensity or projection views~\cite{gao2014low}. This paper focuses on the latter,  \textit{i.e.}, sparse-view CBCT reconstruction. 

Sparse-view CBCT reconstruction aims to retrieve a volumetric attenuation coefficient field from dozens of projections. It is a challenging task in two respects. First, insufficient views lead to notable artifacts. As a comparison, the traditional CBCT obtains hundreds of images. The inputs of sparse-view CBCT are 10$\times$ fewer. Second, the spatial and computational complexity of CBCT reconstruction is much higher than that of FBCT reconstruction due to the dimensional increase of inputs. CBCT relies on 2D projections to build a 3D model, while FBCT simplifies the process by stacking 2D slides restored from 1D projections (but in the sacrifice of time and dose).

Existing CBCT approaches can be divided into three categories: analytical, iterative and learning-based methods. Analytical methods estimate attenuation coefficients by solving the Radon transform and its inverse. A typical example is the FDK algorithm~\cite{feldkamp1984practical}. It produces good results in an ideal scenario but copes poorly with ill-posed problems such as sparse views. The second family, iterative methods, formulates reconstruction as a minimization process. These approaches utilize an optimization framework combined with regularization modules. While iterative methods perform well in ill-posed problems~\cite{andersen1984simultaneous,sidky2008image}, they require substantial computation time and memory. Recently, learning-based methods have become popular with the rise of AI. They use deep neural networks to 1) predict and extrapolate projections~\cite{anirudh2018lose,tang2019projection,wang2021improving,zang2021intratomo}, 2) regress attenuation coefficients with similar data~\cite{kasten2020end,ying2019x2ct}, and 3) make optimization process differentiable~\cite{adler2018learned,chen2017learned,kang2018deep}. Most of these methods~\cite{anirudh2018lose,kasten2020end,tang2019projection,ying2019x2ct} need extensive datasets for network training. Moreover, they rely on neural networks to remember what a CT looks like. Therefore it is difficult to apply a trained model of one application to another. While there are self-supervised methods~\cite{adler2018learned,zang2021intratomo}, they operate under FBCT settings considering network capacity and memory consumption. Their performance and efficiency drop when applied to the CBCT scenario.

Apart from the aforementioned work designated for CT reconstruction, efforts have been made to deal with other ill-posed problems, such as 3D reconstruction in the computer vision field. Similar to CT reconstruction, 3D reconstruction uses RGB images to estimate 3D shapes, which are usually represented as discrete point clouds or meshes. Recent studies propose~\cite{mildenhall2020nerf,park2019deepsdf} Implicit Neural Representation (INR) as an alternative to those discrete representations. INR parameterizes a bounded scene as a neural network that maps spatial coordinates to metrics such as occupancy and color. With the help of position encoder~\cite{mueller2022instant,tancik2020fourier}, INR is capable to learn high-frequency details.

This paper proposes Neural Attenuation Fields (NAF), a fast self-supervised solution for sparse-view CBCT reconstruction. Here we use `self-supervised' to highlight that NAF requires no external CT scans but the X-ray projections of the interested object. Inspired by 3D reconstruction work~\cite{mildenhall2020nerf,park2019deepsdf}, we parameterize the attenuation coefficient field as an INR and imitates the X-ray attenuation process with a self-supervised network pipeline. Specifically, we train a Multi-Layer Perceptron (MLP), whose input is an encoded spatial coordinate $(x,y,z)$ and whose output is the attenuation coefficient $\mu$ at that location. Instead of using a common frequency-domain encoding, we adopt hash encoding~\cite{mueller2022instant}, a learning-based position encoder, to help the network quickly learn high-frequency details. Projections are synthesized by predicting the attenuation coefficients of sampled points along ray trajectories and attenuating incident beams accordingly. The network is optimized with gradient descent by minimizing the error between real and synthesized projections. We demonstrate that NAF quantitatively and qualitatively outperforms existing solutions on both human organ and phantom datasets. While most INR approaches take hours for training, our method can reconstruct a detailed CT model within 10-40 minutes, which is comparable to iterative methods.

In summary, the main contributions of this work are:
\begin{itemize}
    \item We propose a novel and fast self-supervised method for sparse-view CBCT reconstruction. Neither external datasets nor structural prior is needed except projections of a subject.
    \item The proposed method achieves state-of-the-art accuracy and spends relatively short computation time. The performance and efficiency of our method make it feasible for clinical CT applications.
    \item The code will be publicly available for investigation purposes.
\end{itemize}

\begin{figure}[!b]
    %\vspace{-1em}
    \centering
    \includegraphics[width=\linewidth]{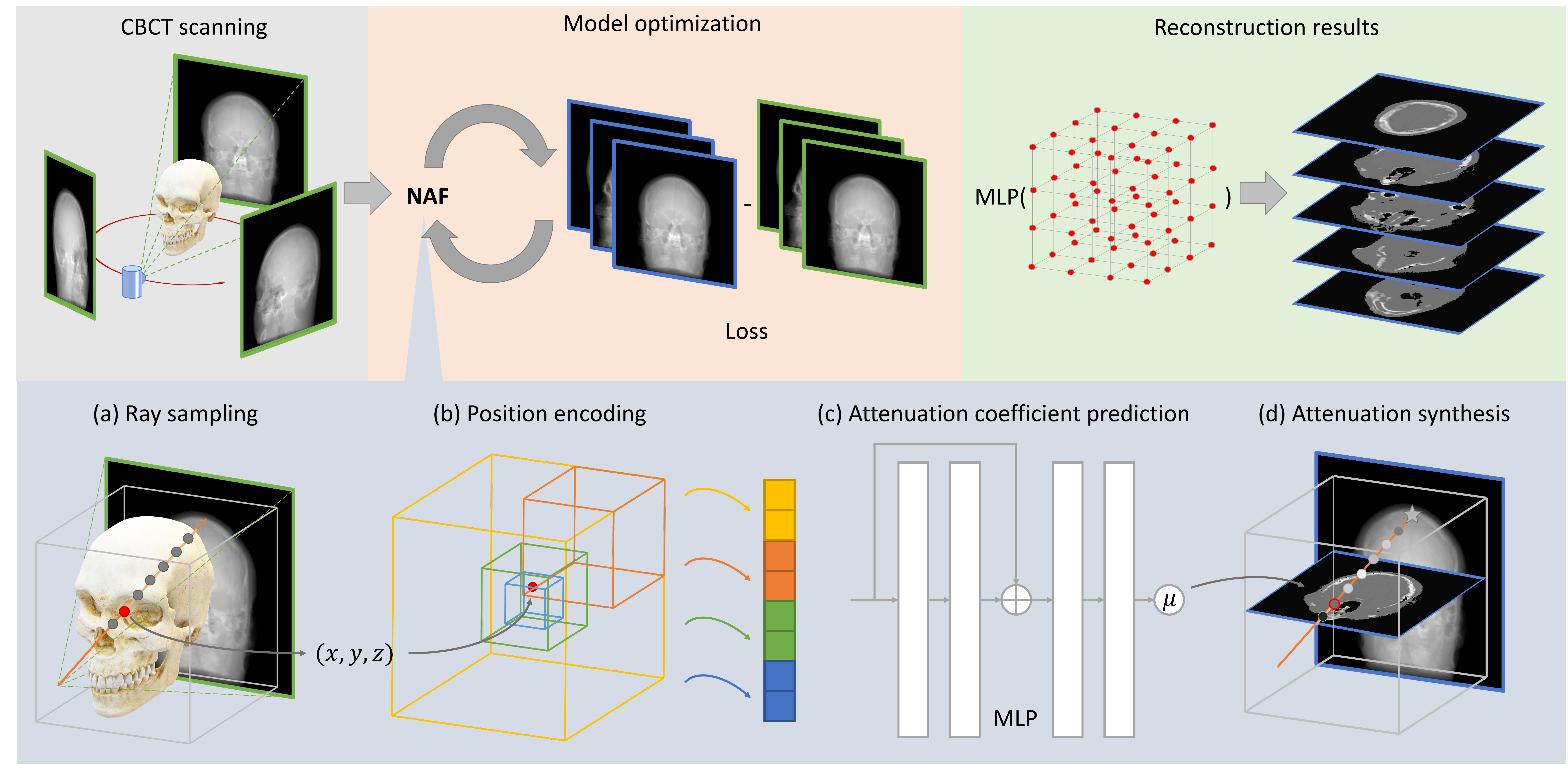}
    \caption{
    % An overview of the proposed 
    NAF pipeline. Gray block: The CBCT scanner captures X-ray projections from different views. Blue block: NAF simulates projections. Orange block: NAF is optimized by comparing real and synthesized projections. Green block: NAF generates a CT model by querying corresponding voxels.}
    \label{fig:framework}
\end{figure}

%%%%%%%%%%%%%%%%%%%%%%%%%
% Method
%%%%%%%%%%%%%%%%%%%%%%%%%
\section{Method}
\subsection{Pipeline}
The pipeline of NAF is shown in Fig.~\ref{fig:framework}. During a CBCT scanning, an X-ray source rotates around the object and emits cone-shaped X-ray beams. A 2D panel detects X-ray projections at equal angular intervals. NAF then uses the scanner geometry to imitate the attenuation process discretely. It learns CT shapes by comparing real and synthesized projections. After the model optimization, the final CT image is generated by querying corresponding voxels.

NAF consists of four modules: ray sampling, position encoding, attenuation coefficient prediction, and projection synthesis. First, we uniformly sample points along X-ray paths based on the scanner geometry. A position encoder network then encodes their spatial coordinates to extract valuable features. After that, an MLP network consumes the encoded information and predicts attenuation coefficients. The last step of NAF is to synthesize projections by attenuating incident X-rays according to the predicted attenuation coefficients on their paths.

\subsection{Neural attenuation fields}
\subsubsection{Ray sampling}
Each pixel value of a projection image results from an X-ray passing through a cubical space and getting attenuated by the media inside. We sample $N$ points at the parts where rays intersect the cube. A stratified sampling method~\cite{mildenhall2020nerf} is adopted, where we divide a ray into $N$ evenly spaced bins and uniformly sample one point at each bin. Setting $N$ greater than the desired CT size ensures that at least one sample is assigned to every grid cell that an X-ray traverses. The coordinates of sampled points are then sent to the position encoding module.

\subsubsection{Position encoding}
A simple MLP can theoretically approximate any function~\cite{hornik1989multilayer}. Recent studies~\cite{rahaman2019spectral,tancik2020fourier}, however, reveal that a neural network prefers to learn low-frequency details due to ``spectral bias''. To this end, a position encoder is introduced to map 3D spatial coordinates to a higher dimensional space.

A common choice is the \textit{frequency encoder} proposed by Mildenhall \textit{et al.}~\cite{mildenhall2020nerf}. It decomposes a spatial coordinate $\mathbf{p}\in\mathbb{R}^{3}$ into $L$ sets of sinusoidal components at different frequencies. While frequency encoder eases the difficulty of training networks, it is considered quite cumbersome. In medical imaging practise~\cite{wu2021irem,zang2021intratomo}, the size of encoder output is set to 256 or greater. The following network must be wider and deeper to cope with the inflated inputs. As a result, it takes hours to train millions of network parameters, which is not acceptable for fast CT reconstruction.

Frequency-domain encoding is a dense encoder because it utilizes the entire frequency spectrum. However, dense encoding is redundant for CBCT reconstruction for two main reasons. First, a  human body usually consists of several homogeneous media, such as muscles and bones. Attenuation coefficients remain approximately uniform inside one medium but vary between different media. High-frequency features are not necessary unless for points near edges. Second, natural objects favor smoothness. Many organs have simple shapes, such as spindle (muscle) or cylinder (bone). Their smooth surfaces can be easily learned with low-dimensional features. 

To exploit the aforementioned characteristics of the scanned objects, we use the \textit{hash encoder}~\cite{mueller2022instant}, a learning-based sparse encoding solution. The equation of hash encoder $\mathcal{M_{H}}$ is: 
\begin{equation}
    \mathcal{M_{H}}(\mathbf{p};\mathbf{\Theta})=[\mathcal{I}(\mathbf{H}_{1}),\cdots,\mathcal{I}(\mathbf{H}_{L})]^T,~\mathbf{H}=\{\mathbf{c}|h(\mathbf{c})=(\bigoplus c_{j}\pi_{j})~{\rm mod}~T\}.
    \end{equation}
Hash encoder describes a bounded space by $L$ multiresolution voxel grids. A trainable feature lookup table $\mathbf{\Theta}$ with size $T$ is assigned to each voxel grid. At each resolution level, we 1) detect neighbouring corners $\mathbf{c}$  (cubes with different colors in Fig.~\ref{fig:framework}(b)) of the queried point $\mathbf{p}$, 2) look up their corresponding features $\mathbf{H}$ in a hash function fashion $h$~\cite{teschner2003optimized}, and 3) generate a feature vector with linear interpolation $\mathcal{I}$. The output of a hash encoder is the concatenation of feature vectors at all resolution levels. More details of hash function and its symbols can be found in~\cite{mueller2022instant}.

Compared with frequency encoder, hash encoder produces much smaller outputs ($32$ in our setting) with competitive feature quality for two reasons. On the one hand, the many-to-one property of hash function conforms to the sparsity nature of human organs. On the other hand, a trainable encoder can learn to focus on relevant details and select suitable frequency spectrum~\cite{mueller2022instant}. Thanks to hash encoder, the subsequent network 
is more compact.

\subsubsection{Attenuation coefficient prediction}
We represent the bounded field with a simple MLP $\mathbf{\Phi}$, which takes the encoded spatial coordinates as inputs and outputs the attenuation coefficients $\mu$ at that position. As illustrated in Fig.~\ref{fig:framework}(c), the network is composed of 4 fully-connected layers. The first three layers are 32-channel wide and have ReLU activation functions in between, while the last layer has one neuron followed by a sigmoid activation. A skip connection is included to concatenate the network input to the second layer's activation. By contrast, Zang \textit{et al.}~\cite{zang2021intratomo} use a 6-layer 256-channel MLP to learn features from a frequency encoder. Our network is $10\times$ smaller.

\subsubsection{Attenuation synthesis}
According to Beer's Law, the intensity of an X-ray traversing matter is reduced by the exponential integration of attenuation coefficients on its path. We numerically synthesize the attenuation process with:

\begin{equation}
    I=I_{0}\exp(-\sum_{i=1}^{N}\mu_{i}\delta_{i}),
\end{equation}
where $I_{0}$ is the initial intensity and $\delta_{i}=\|\mathbf{p}_{i+1}-\mathbf{p}_{i}\|$ is the distance between adjacent points.

\subsection{Model optimization and output}
NAF is updated by minimizing the L2 loss between real and synthesized projections. The loss function $\mathcal{L}$ is defined as:
\begin{equation}
   \mathcal{L}(\mathbf{\Theta},\mathbf{\Phi}) = \sum_{\mathbf{r}\in\mathbf{B}}\|I_{r}(\mathbf{r})-I_{s}(\mathbf{r})\|^2,
\end{equation}
where $\mathbf{B}$ is a ray batch, and $I_{r}$ and $I_{s}$ are real and synthesized projections for ray $\mathbf{r}$ respectively. We update both hash encoder $\mathbf{\Theta}$ and attenuation coefficient network $\mathbf{\Phi}$ during the training process.

The final output is formulated as a discrete 3D matrix. We build a voxel grid with the desired size and pass the voxel coordinates to the trained MLP to predict the corresponding attenuation coefficients. A CT model thus is restored.

%%%%%%%%%%%%%%%%%%%%%%%%%
% Experiments
%%%%%%%%%%%%%%%%%%%%%%%%%
\section{Experiments}
\subsection{Experimental settings}

\subsubsection{Data}
We conduct experiments on five datasets containing human organ and phantom data. Details are listed in Table~\ref{tab:dataset}. 

% \subsubsection{Human organ} 
\noindent{\textit{\textbf{Human organ}}}: We evaluate our method using public datasets of human organ CTs~\cite{armato2011lung,Klacansky2022Open}, including chest, jaw, foot and abdomen. The chest data are from LIDC-IDRI dataset~\cite{armato2011lung}, and the rest are from Open Scientific Visualization Datasets~\cite{Klacansky2022Open}. Since these datasets only provide volumetric CT scans, we generate projections by a tomographic toolbox TIGRE~\cite{biguri2016tigre}. In TIGRE~\cite{biguri2016tigre}, we capture 50 projections with 3\% noise in the range of 180\degree. We train our model with these projections and evaluate its performance with the raw volumetric CT data.

\noindent{\textit{\textbf{Phantom}}}: We collect a phantom dataset by scanning a silicon aortic phantom with GE C-arm Medical System. This system captures 582 500$\times$500 fluoroscopy projections with position primary angle from -103\degree to 93\degree and position secondary angle of 0\degree. A 512$\times$512$\times$510 CT image is also generated with inbuilt algorithms as the ground truth. We only use 50 projections for experiments.

\begin{table}[!b]
\scriptsize
\vspace{-2em}
\centering
\caption{Details of CT datasets used in the experiments.}
\label{tab:dataset}
\begin{tabular}{m{0.17\columnwidth}<{\centering}|m{0.15\columnwidth}<{\centering}m{0.15\columnwidth}<{\centering}m{0.15\columnwidth}<{\centering}m{0.15\columnwidth}<{\centering}m{0.15\columnwidth}<{\centering}}
\toprule
Dataset name & CT dimension & Scanning method & Scanning range & Number of projections & Detector resolution \\ \midrule
Chest~\cite{armato2011lung}   & 128$\times$128$\times$128  & TIGRE~\cite{biguri2016tigre} & $0\degree\sim180\degree$ & 50 & 256$\times$256 \\
Jaw~\cite{Klacansky2022Open}     & 256$\times$256$\times$256  & TIGRE~\cite{biguri2016tigre} & $0\degree\sim180\degree$ & 50 & 512$\times$512 \\
Foot~\cite{Klacansky2022Open}    & 256$\times$256$\times$256  & TIGRE~\cite{biguri2016tigre} & $0\degree\sim180\degree$ & 50 & 512$\times$512 \\
Abdomen~\cite{Klacansky2022Open} & 512$\times$512$\times$463  & TIGRE~\cite{biguri2016tigre} & $0\degree\sim180\degree$ & 50 & 1024$\times$1024 \\
Aorta   & 512$\times$512$\times$510  & GE C-arm & $-103\degree\sim93\degree$ & 50 (582) & 500$\times$500 \\ \bottomrule
\end{tabular}
\end{table}

\subsubsection{Baselines}
We compare our approach with four baseline techniques. \textbf{FDK}~\cite{feldkamp1984practical} is firstly chosen as a representative of analytical methods. The second method \textbf{SART}~\cite{andersen1984simultaneous} is a robust iterative reconstruction algorithm. \textbf{ASD-POCS}~\cite{sidky2008image} is another iterative method with a total-variation regularizer. We implement a CBCT variant of IntraTomo~\cite{zang2021intratomo}, named  \textbf{IntraTomo3D}, as an example of frequency-encoding deep learning methods.

\subsubsection{Implementation details}
Our proposed method is implemented in PyTorch~\cite{NEURIPS2019_9015}. We use Adam optimizer with a learning rate that starts at $1\times10^{-3}$ and steps down to $1\times10^{-4}$. The batch size is 2048 rays at each iteration. The sampling quantity of each ray depends on the size of CT data. For example, we sample $192$ points along each ray for the 128$\times$128$\times$128 chest CT. We use the same hyper-parameter setting for hash encoder as~\cite{mueller2022instant}. More details of hyper-parameters can be found in the supplementary material. All experiments are conducted on a single RTX 3090 GPU. We evaluate five methods quantitatively in terms of peak signal-to-noise ratio (PNSR) and structural similarity (SSIM)~\cite{wang2004image}. PSNR (dB) statistically assesses the artifact suppression performance, while SSIM measures the perceptual difference between two signals. Higher PNSR/SSIM values represent the accurate reconstruction and vice versa.

\subsection{Results}
\subsubsection{Performance} 
Our method produces quantitatively best results in both human organ and phantom datasets as listed in Table~\ref{tab:results}. Both PSNR and SSIM values are significantly higher than other methods. For example, the PSNR value of our method in the abdomen dataset is 3.07 dB higher than that of the second-best method \textbf{SART}. 

\begin{figure}[h]
    \centering
    \includegraphics[width=\linewidth]{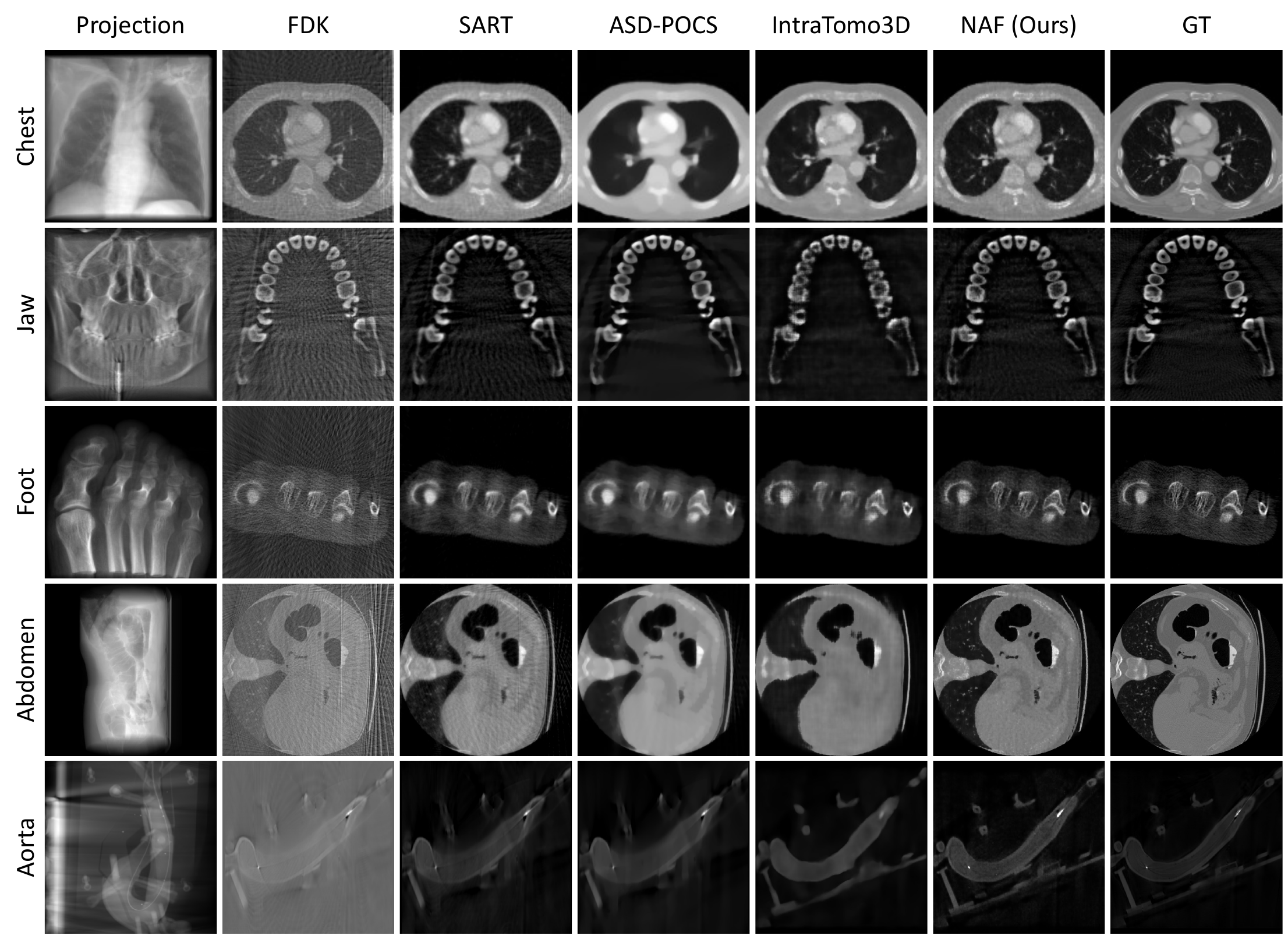}
    \caption{Qualitative results of five methods. From left to right: examples of X-ray projections, slices of 3D CT models reconstructed by five methods, and the ground truth CT slices.}
    \label{fig:demo}
\end{figure}

\begin{figure}[t]
\centering
\begin{minipage}[t]{0.47\linewidth}
\centering
\includegraphics[width=\linewidth]{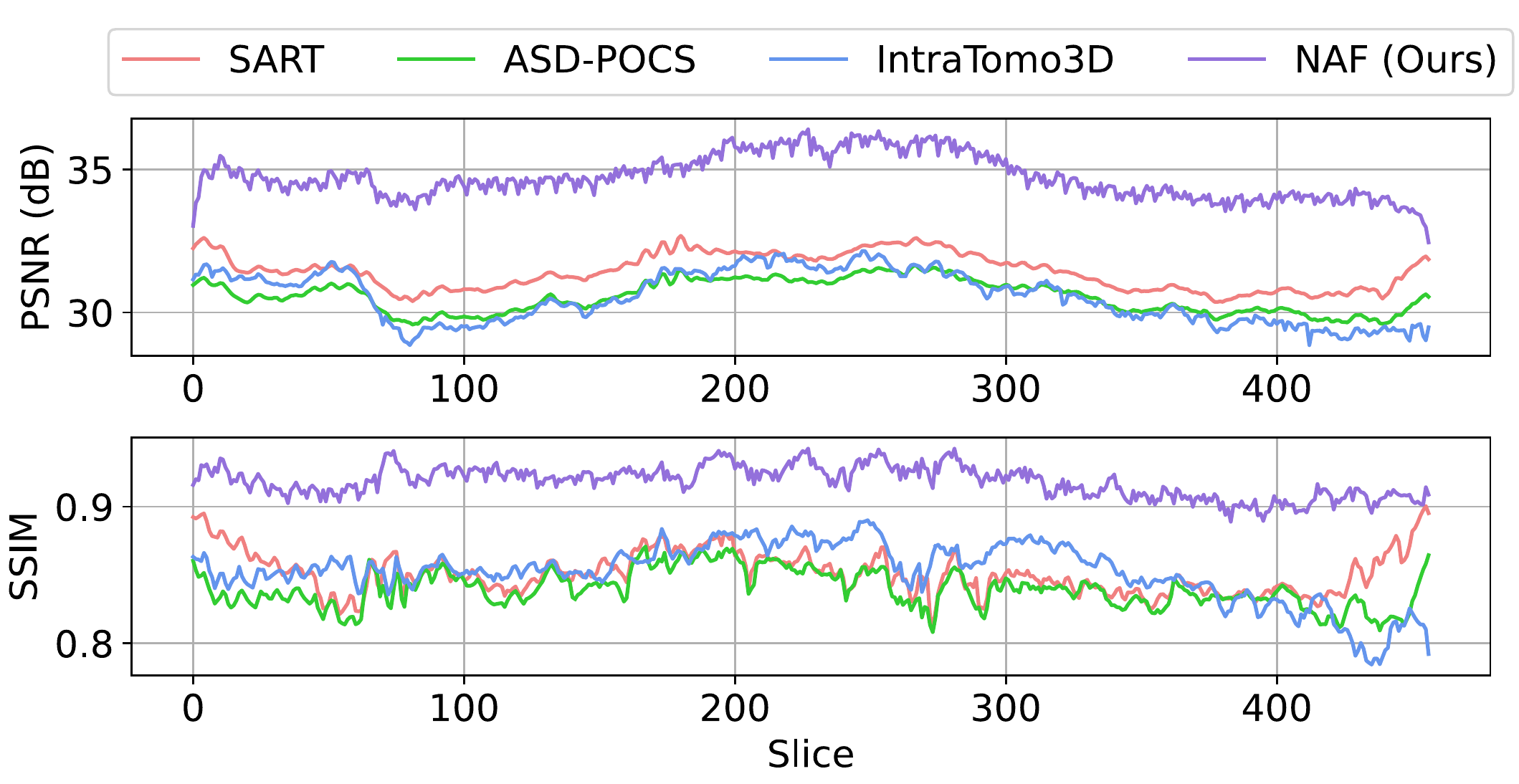}
\caption{Slice-wise performance of iterative and learning-based methods on the abdomen dataset.}
 \label{fig:slice}
\end{minipage}
\hspace{4mm}
\begin{minipage}[t]{0.47\linewidth}
\centering
\includegraphics[width=\linewidth]{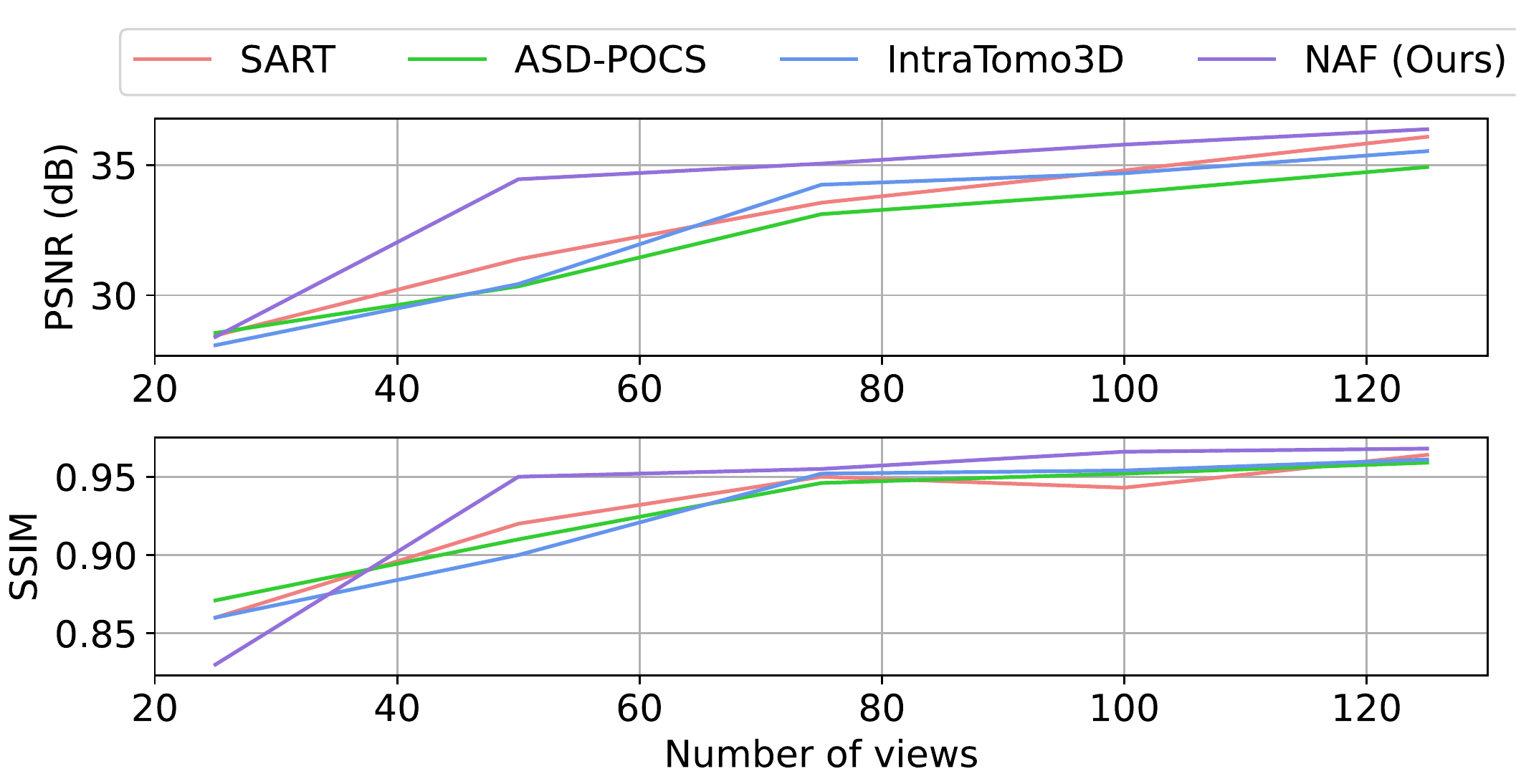}
\caption{Performance under different number of views on the abdomen dataset.}
 \label{fig:numView}
\end{minipage}
\vspace{-1em}
\end{figure}

We also provide visualization results of different methods in Fig.~\ref{fig:demo}. \textbf{FDK} restores low-quality models with notable artifacts, as analytical methods demand large amounts of projections.
\begin{table}[h]
\vspace{-1.5em}
\caption{PSNR/SSIM measurements of five methods on five datasets.}
\label{tab:results}
\centering
\begin{tabular}{m{0.25\columnwidth}<{\centering}|m{0.14\columnwidth}<{\centering}m{0.14\columnwidth}<{\centering}m{0.14\columnwidth}<{\centering}m{0.14\columnwidth}<{\centering}m{0.14\columnwidth}<{\centering}}
\toprule
            & Chest         & Jaw & Foot & Abdomen & Aorta \\ \midrule
\textbf{FDK}~\cite{feldkamp1984practical}         & 22.89/.78    & 28.59/.78   &  23.92/.58    & 22.39/.59        &  12.11/.21     \\
\textbf{SART}~\cite{andersen1984simultaneous}       & 32.12/.95    & 32.67/.93     & 30.13/.93      & 31.38/.92         &  27.31/.77     \\
\textbf{ASD-POCS}~\cite{sidky2008image}    & 29.78/.92    & 32.78/.93   &  28.67/.89       & 30.34/.91     &  27.30/.76     \\
\textbf{IntraTomo3D}~\cite{zang2021intratomo}   & 31.94/.95    & 31.95/.91  & 31.43/.91      &  30.43/.90       &  29.38/.82     \\
\textbf{NAF (Ours)} & \textbf{33.05}/\textbf{.96}    & \textbf{34.14}/\textbf{.94}     & \textbf{31.63}/\textbf{.94}     & \textbf{34.45}/\textbf{.95}        &  \textbf{30.34}/\textbf{.88}     \\ 
\bottomrule
\end{tabular}
\vspace{-3em}
\end{table}
Iterative method \textbf{SART} suppresses noise in the sacrifice of losing certain details. The reconstruction results of \textbf{ASD-POCS} are heavily smeared because total-variation regularization encourages removing high-frequency details, including unwanted noise and expected tiny structures. \textbf{IntraTomo3D} produces clean results. However, edges between media are slightly blurred, which shows that the frequency encoder fails to teach the network to focus on edges. With the help of hash encoding, results of the proposed \textbf{NAF} have the most details, clearest edges and fewest artifacts. Fig.~\ref{fig:slice} indicates that \textbf{NAF} outperforms other methods in all slices of the reconstructed CT volume. 

Figure~\ref{fig:numView} shows the performance of iterative methods and learning-based methods under different number of views. It is clear that the performance increases with the rise of input views. Our methods achieves better results than others under most circumstances.

\subsubsection{Time} 
\begin{figure}[t]
    \centering
    \includegraphics[width=\linewidth]{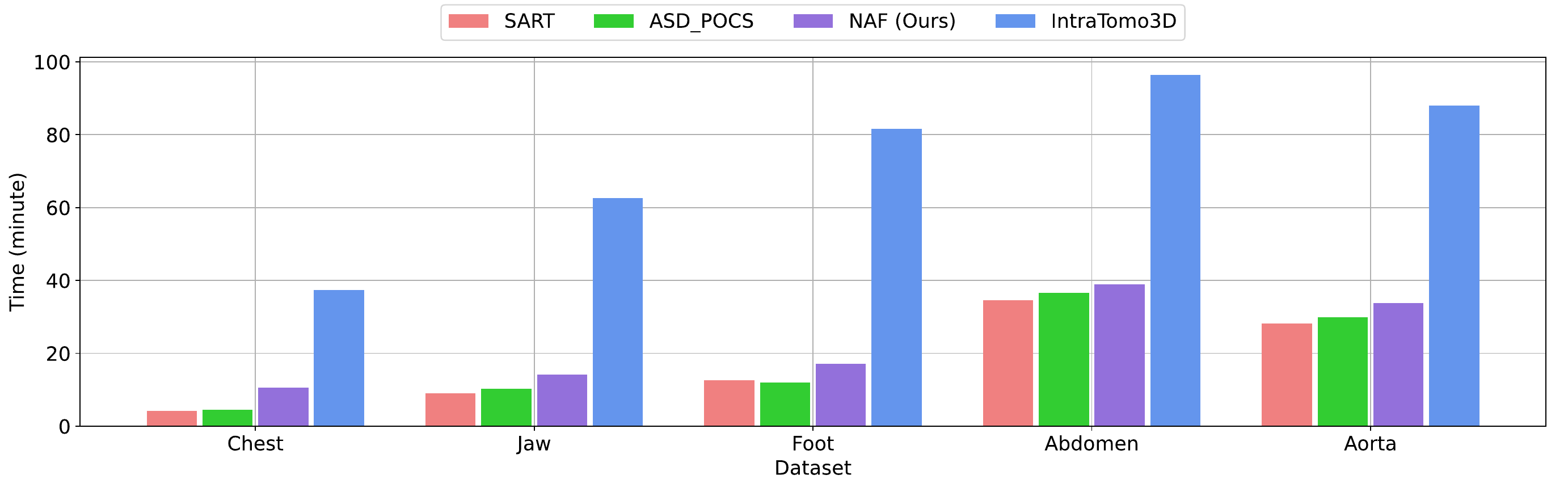}
    \caption{Running time that iterative and learn-based methods take to converge to stable results.}
    \label{fig:time}
\end{figure}

We record the running time of iterative and learning-based methods as shown in Fig.~\ref{fig:time}. All methods use CUDA~\cite{cuda} to accelerate the computation process. Overall, the methods spend less time on datasets with small projections (chest, jaw and foot) and increasingly more time on big datasets (abdomen and aorta). \textbf{IntraTomo3D} requires more than one hour to train the network. Benefiting from the compact network design, \textbf{NAF} spends similar running time to iterative methods and is 3$\times$ faster than the frequency-encoding deep learning method \textbf{IntraTomo3D}.

%%%%%%%%%%%%%%%%%%%%%%%%%
% Conclusion
%%%%%%%%%%%%%%%%%%%%%%%%%
\section{Conclusion}
This paper proposes NAF, a fast self-supervised learning-based solution for sparse-view CBCT reconstruction. Our method trains a fully-connected deep neural network that consumes a 3D spatial coordinate and outputs the attenuation coefficient at that location. NAF synthesizes projections by attenuating incident X-rays based on the predicted attenuation coefficients. The network is updated by minimizing the projection error. We show that frequency encoding is not computationally efficient for tomographic reconstruction tasks. As an alternative, a learning-based encoder entitled hash encoding is adopted to extract valuable features. Experimental results on human organ and phantom datasets indicate that the proposed method achieves significantly better results than other baselines and spends reasonably short computation time.

% \begin{figure}[htbp]
%     \centering
%     \subfigure[Running time.]{  
%     \begin{minipage}[t]{0.45\linewidth}
%     \centering 
%     \includegraphics[width=\linewidth]{figures/time.pdf}
%     \label{fig:time}
%     \end{minipage}
%     }
%     \subfigure[Abdomen CT results.]{
%     \begin{minipage}[t]{0.45\linewidth}
%     \centering  
%     \includegraphics[width=\linewidth]{figures/slice_abdomen_50.pdf}
%     \label{fig:slice}
%     \end{minipage}
%     }
%     \caption{Left: Running time of five methods. Right: PSNR/SSIM measurements of the reconstructed abdomen CT slices.}  
%     \label{fig:demo2}   
% \end{figure}

%
% ---- Bibliography ----
%
% BibTeX users should specify bibliography style 'splncs04'.
% References will then be sorted and formatted in the correct style.
%
\bibliographystyle{splncs04}
\bibliography{main}

\end{document}